\documentstyle[11pt]{article}

\begin{document}

\begin{flushright}
DPNU-01-10
\end{flushright}
\vspace{-0.7cm}
\begin{center}
\Large\bf
Chiral Perturbation in the Hidden Local Symmetry and
\\
Vector Manifestation of Chiral Symmetry
\footnote{%
Talk presented at 
8th International Symposium on Particle Strings and Cosmology 
(PASCOS 2001), Chapel Hill, North Carolina, 10-15 Apr 2001. 
This talk is based on the works done in collaboration with
Prof. Yamawaki~\cite{HY:letter, HY:matching, HY:VM}.}
\end{center}

\begin{center}
{\Large Masayasu Harada}\\
{\it
Department of Physics, Nagoya University, Nagoya 464-8602, JAPAN
}
\end{center}

\begin{abstract}
In this talk I summarize our recent
works on the chiral phase transition in the large flavor
QCD studied by the hidden local symmetry (HLS).
Bare parameters in the HLS are determined by matching the HLS
with the underlying QCD at the matching scale through the Wilsonian
matching.
This leads
to the vector manifestation of the Wigner realization
of the chiral symmetry in
which the symmetry is restored 
by the massless degenerate pion (and its flavor
partners) and rho meson (and its flavor partners) as the chiral
partner.
\end{abstract}

\nocite{HY:letter}\nocite{HY:letter}
\nocite{HY:matching}
\nocite{HY:VM}

Chiral phase transition in QCD is discussed in various contexts such
as the large flavor QCD and the hot and/or dense QCD, etc.
Recently, in Refs.~\ref{ref:HY:letter}, \ref{ref:HY:matching} and
\ref{ref:HY:VM}, 
we studied the chiral phase transition 
in the large flavor QCD using the hidden local
symmetry (HLS) model~\cite{BKUYY},
which is an effective field theory of QCD
including vector and pseudoscalar mesons.

The chiral symmetry restoration in the large $N_f$ QCD 
($< \frac{11}{2} N_c$) was implied by the fact that the coupling at
the infrared fixed point becomes very small~\cite{BZ}.
Such a restoration was indeed observed by various methods like lattice
simulation~\cite{lattice}, ladder Schwinger-Dyson
equation~\cite{ATW,MY}, dispersion relation~\cite{OZ},
instanton calculus~\cite{VS}, etc.

In Ref.~\ref{ref:HY:letter}
we pointed that the chiral restoration takes place for large $N_f$
also in the HLS {\it by its own dynamics}.
Inclusion of the quadratic divergences in the renormalization group
equations (RGE's) was essential 
to obtain the phase transition.
Here I should emphasize that 
{\it thanks to the gauge symmetry}
in the HLS it is possible to perform 
a {\it systematic loop expansion}
including the vector mesons in addition to the pseudoscalar 
mesons~\cite{Georgi,Tanabashi,HY,LET,HY:matching}
in a way to extend the chiral perturbation theory~\cite{Wei:79,GL}.
There the loop expansion corresponds to the derivative
expansion, so that the one-loop calculation of the RGE is reliable in
the low-energy region.
The quadratic divergence in the RGE's
yields the quadratic running of (square of)
the decay constant $F_\pi^2(\mu)$, where $\mu$ is the renormalization
point.  What is shown in Ref.~\ref{ref:HY:letter} is that 
{\it the order
parameter $F_\pi(0)$ can become zero for larger $N_f$ even when
$F_\pi(\Lambda) \neq 0$}, where $F_\pi(\Lambda)$ is not the order
parameter but just a parameter of the bare Lagrangian defined at the
cutoff $\Lambda$ where the matching with QCD is made.

In Ref.~\ref{ref:HY:matching}
we proposed a novel way of matching the HLS
with the underlying QCD in the sense of a Wilsonian RGE,
namely, including quadratic divergences in the HLS (``Wilsonian
matching'').
The basic tool of the Wilsonian matching is the Operator Product
Expansion (OPE) of QCD for the axialvector and vector current
correlators, which are equated with those from the HLS at the
matching scale $\Lambda$.  
{\it This determines without much ambiguity the
bare parameters of the HLS defined at the scale $\Lambda$ in terms of
the QCD parameters}.
It is shown that the physical quantities for the $\pi$ and $\rho$
system are calculated by the Wilsonian RGE's from the bare parameters
in remarkable agreement with experiment.

In Ref.~\ref{ref:HY:VM} we
applied the Wilsonian matching for the large $N_f$ QCD,
and proposed ``Vector Manifestation'' (VM) of the chiral symmetry
as a novel manifestation of the Wigner realization
in which 
{\it the vector meson denoted by $\rho$
($\rho$ meson and its flavor partner)
becomes massless at the chiral
phase transition point}.
Accordingly, {\it the (longitudinal) $\rho$
becomes the chiral partner of the Nambu-Goldstone (NG) boson
denoted by $\pi$ (pion and its flavor partners)}.

Below I shall summarize main points of the Wilsonian matching
and the VM.

Let me start from 
the Wilsonian matching~\cite{HY:matching}
of the HLS with the underlying QCD.
In the HLS 
axialvector and vector current correlators are
well described by the tree contributions with including
${\cal O}(p^4)$ terms
when the momentum is around the matching scale, $Q^2 \sim \Lambda^2$:
\begin{eqnarray}
\Pi_A^{\rm(HLS)}(Q^2) 
=
\frac{F_\pi^2(\Lambda)}{Q^2} - 2 z_2(\Lambda)
\ ,
\quad
\Pi_V^{\rm(HLS)}(Q^2) 
=
\frac{
  F_\sigma^2(\Lambda)
  \left[ 1 - 2 g^2(\Lambda) z_3(\Lambda) \right]
}{
  M_v^2(\Lambda) + Q^2
} 
- 2 z_1(\Lambda)
\ ,
\label{Pi A V HLS}
\end{eqnarray}
where $g(\Lambda)$ is the bare HLS gauge coupling,
$F_\sigma^2(\Lambda) = a(\Lambda) F_\pi^2(\Lambda)$ the bare decay
constant of the would-be NG boson $\sigma$ 
absorbed into the HLS gauge boson $\rho$, and 
$M_v^2(\Lambda) \equiv g^2(\Lambda) F_\sigma^2(\Lambda)$ the bare
$\rho$ mass.
$z_1(\Lambda)$, $z_2(\Lambda)$ and $z_3(\Lambda)$ are the (bare)
coefficients of the relevant ${\cal O}(p^4)$ terms.
The same correlators are evaluated by the OPE up until 
${\cal O}(1/Q^6)$~\cite{SVZ}:
\begin{eqnarray}
&&
\Pi_A^{\rm(QCD)}(Q^2) = \frac{1}{8\pi^2}
\Biggl[
  - \left( 1 + \frac{\alpha_s}{\pi} \right) \ln \frac{Q^2}{\mu^2}
  + \frac{\pi^2}{3} 
    \frac{
      \left\langle 
        \frac{\alpha_s}{\pi} G_{\mu\nu} G^{\mu\nu}
      \right\rangle
    }{ Q^4 }
  + \frac{\pi^3}{3} \frac{1408}{27}
    \frac{\alpha_s \left\langle \bar{q} q \right\rangle^2}{Q^6}
\Biggr]
\ ,
\nonumber\\
&&
\Pi_V^{\rm(QCD)}(Q^2) = \frac{1}{8\pi^2}
\Biggl[
  - \left( 1 + \frac{\alpha_s}{\pi} \right) \ln \frac{Q^2}{\mu^2}
  + \frac{\pi^2}{3} 
    \frac{
      \left\langle 
        \frac{\alpha_s}{\pi} G_{\mu\nu} G^{\mu\nu}
      \right\rangle
    }{ Q^4 }
  - \frac{\pi^3}{3} \frac{896}{27}
    \frac{\alpha_s \left\langle \bar{q} q \right\rangle^2}{Q^6}
\Biggr]
\ ,
\label{Pi A V OPE}
\end{eqnarray}
where $\mu$ is the renormalization scale of QCD.
The current correlators in the HLS
in Eq.~(\ref{Pi A V HLS}) can be 
matched with those in QCD in Eq.~(\ref{Pi A V OPE}) up until first
derivatives at $\Lambda$:
\begin{equation}
\Pi_{A}^{\rm (HLS)} - \Pi_{V}^{\rm (HLS)}
=
\Pi_{A}^{\rm (QCD)} - \Pi_{V}^{\rm (QCD)}
\ ,
\quad
\frac{d}{dQ^2} \Pi_{A,V}^{\rm (HLS)}
=
\frac{d}{dQ^2} \Pi_{A,V}^{\rm (QCD)}
\ ,
\quad
\mbox{at} \ Q^2 = \Lambda^2 \ .
\label{WM cond}
\end{equation}
Here the difference of two current correlators in the first equation 
is taken to eliminate the explicit dependence on the renormalization
scale $\mu$ of QCD.
The above three equations
are the Wilsonian matching conditions proposed in
Ref.~\ref{ref:HY:matching}.

The right-hand sides in three equations in (\ref{WM cond})
are directly determined from QCD.
First note that the matching scale $\Lambda$ must be small enough for
the validity of the systematic expansion in the HLS,
whereas $\Lambda$ has to be big enough for the OPE to be valid.
Here I take $\Lambda = 1.1\,\mbox{GeV}$.
To determine the current correlators from the OPE we used
\begin{equation}
\left\langle \frac{\alpha_s}{\pi} G_{\mu\nu} G^{\mu\nu}
\right\rangle = 0.012 \,\mbox{GeV}^4 \ ,
\quad
\left\langle \bar{q} q \right\rangle_{\rm 1\,GeV} = - 
\left(\mbox{0.25\,GeV}\right)^3 \ ,
\end{equation}
shown in Ref.~\ref{ref:SVZ}\nocite{SVZ}
and 
$\Lambda_{\rm QCD} = 400\,\mbox{MeV}$
as typical values.
One-loop running is used to
estimate $\alpha_s(\Lambda)$ and 
$\left\langle \bar{q} q \right\rangle_\Lambda$.
Then the bare parameters $F_\pi(\Lambda)$, $a(\Lambda)$,
$g(\Lambda)$, $z_3(\Lambda)$ and $z_2(\Lambda)-z_1(\Lambda)$
can be determined
through the Wilsonian matching conditions.
Actually, the Wilsonian matching
conditions in Eq.~(\ref{WM cond})
are not enough to determine all the relevant bare
parameters.  Therefore, the on-shell pion decay constant
$F_\pi(0)=88$\,MeV in the chiral limit~\cite{GL} and the
$\rho$ mass $m_\rho = 770$\,MeV are used as inputs.
The resultant values of all the relevant bare parameters of the HLS
are shown in Table~\ref{tab:bare} together with those at
$\mu=m_\rho$.
\begin{table}[htbp]
\begin{center}
\begin{tabular}{|c||c|c|c|c|c|}
\hline
$\mu$ & $F_\pi(\mu)$ & $a(\mu)$ & $g(\mu)$ & $z_3(\mu)$ 
  & $z_2(\mu)-z_1(\mu)$ \\
\hline
$\Lambda=1.1$\,GeV & 0.149 & 1.19 & 3.69 & -5.23$\times10^{-3}$ 
  & -1.03$\times10^{-3}$ \\
$m_\rho=0.77$\,GeV  & 0.110 & 1.22 & 6.33 & -6.34$\times10^{-3}$ 
  & -1.24$\times10^{-3}$ \\
\hline
\end{tabular}
\caption[]{%
Five parameters of the HLS at $\mu=\Lambda$ and $m_\rho$ for 
$\Lambda_{\rm QCD} = 400$\,MeV.
The unit of $F_\pi$ is GeV.
}\label{tab:bare}
\end{center}
\end{table}

Now that the bare Lagrangian has been
completely specified, several physical quantities
can be predicted by the Wilsonian RGE's.
Table~\ref{tab:pred} shows 
two examples of the physical predictions given
in Ref.~\ref{ref:HY:matching}.
\begin{table}[htbp]
\begin{center}
\begin{tabular}{|c||c|c|}
\hline
  & $\Gamma(\rho\rightarrow\pi\pi)$ 
  & $\Gamma(\rho\rightarrow e^+e^-)$ \\
\hline
theory & $151$\,MeV & $6.8$,keV \\
Exp.   & $(150.8\pm2.0)$\,MeV & $(6.77\pm0.32)$\,keV \\
\hline
\end{tabular}
\end{center}
\caption[]{%
Predictions of
$\Gamma(\rho\rightarrow\pi\pi)$ and
$\Gamma(\rho\rightarrow e^+e^-)$
from the Wilsonian matching.%
}\label{tab:pred}
\end{table}
The agreement between the predicted and experimental values 
are excellent.
Especially, the prediction of $\Gamma(\rho\rightarrow e^+e^-)$
is substantially improved from that at the leading order,
$\Gamma(\rho\rightarrow e^+e^-) = 5.2$\,keV. (See, e.g. 
Ref.~\ref{ref:HSc}, \nocite{HSc}
where a tree-level analysis of 
the equivalent model~\cite{KS} was made.)
This shows that the Wilsonian matching works very well for 
the $N_f=3$ QCD.

Now, let me briefly summarize the VM proposed in
Ref.~\ref{ref:HY:VM}.
One important result of the Wilsonian matching 
is that
{\it $F_\pi^2(\Lambda)$ is non-zero even at the critical point}
where the chiral symmetry
is restored with $\left\langle \bar{q} q \right\rangle = 0$.
Then how do we know by the bare parameters defined at $\Lambda$
whether or not the chiral symmetry is restored ?
A clue comes from the fact that
$\Pi_A^{\rm (QCD)}$ 
and $\Pi_V^{\rm (QCD)}$ in Eq.~(\ref{Pi A V OPE})
agree with each other for any value of $Q^2$ at the critical point.
Thus, through the Wilsonian matching,
it is reasonable to require that
$\Pi_A^{\rm (HLS)}$ and $\Pi_V^{\rm (HLS)}$ in Eq.~(\ref{Pi A V HLS})
agree with each other for {\it any value of $Q^2$}.
This agreement is satisfied by the following conditions:
\begin{equation}
g(\Lambda) \rightarrow 0 \ , 
\quad a(\Lambda) = F_\sigma^2(\Lambda)/F_\pi^2(\Lambda)
  \rightarrow 1 \ , \quad
z_1(\Lambda) - z_2(\Lambda) \rightarrow 0 \ .
\label{vector conditions}
\end{equation}

The low-energy phenomena are studied by solving the RGE's.
We found that the on-shell pion decay constant $F_\pi(0)$, which is
actually the order parameter, 
vanishes even though it at the matching scale
$F_\pi(\Lambda)$ is non-zero.
The critical value of $N_f$ is expressed by the
parameters in the OPE as
\begin{equation}
N_f^{\rm cr} = 4 
\left[
  1 + \frac{\alpha_s}{\pi}
  + \frac{2\pi^2}{3} 
    \frac{
      \left\langle 
        \frac{\alpha_s}{\pi} G_{\mu\nu} G^{\mu\nu}
      \right\rangle
    }{ \Lambda^4 }
\right]
\ .
\end{equation}
For wide range of the values of the parameters this is estimated as
$N_f^{\rm cr} \simeq 5$.
Furthermore, since $(g,a)=(0,1)$ is the fixed point of the 
RGE's~\cite{HY:letter},
$g(m_\rho)=0$ and $a(m_\rho)=1$
at the $\rho$ mass scale $m_\rho$.
This implies that
$\rho$ becomes massless ($m_\rho = g(m_\rho)F_\sigma(m_\rho)=0$)
with the current coupling equal to
that of $\pi$ ($F_\sigma(m_\rho) = F_\pi(0)$):
\begin{equation}
m_\rho \rightarrow 0 \ , \quad
F_\sigma(m_\rho) \rightarrow F_\pi(0) \quad
\mbox{for} \quad N_f \rightarrow N_f^{\rm cr} \ .
\end{equation}
This is nothing but the VM of the chiral symmetry
which is accompanied by the
degenerate 
massless $\pi$ and (longitudinal) $\rho$.
A salient feature of the VM is that
{\it $m_\rho$ approaches to zero faster than
$F_\pi$};~\cite{HY:VM}
\begin{equation}
m_\rho^2/F_\pi^2(0) \rightarrow a(m_\rho) g^2(m_\rho) \rightarrow 0
\ .
\end{equation}

Finally, let me make a comment on the application of
the VM to other chiral phase transitions such as the
one at finite temperature and/or density~\cite{BR}.
In such a case, the
position of the $\rho$ peak of the dilepton spectrum would move to the
lower energy region in accord with the picture shown in
Ref.~\ref{ref:BR}.
Furthermore, the VM would imply smaller $\rho$ width 
($\Gamma/m_\rho \sim g_{\rho\pi\pi}^2 \rightarrow 0$)
near the critical point.  If it is really the
case, this would  be clear signals of VM tested in the future
experiments.

\section*{Acknowledgment}

I would like to thank Professor Koichi Yamawaki for collaboration in
Refs.~\ref{ref:HY:letter}, \ref{ref:HY:matching} and \ref{ref:HY:VM}
on which this talk is based.
I would be grateful to the organizers for giving me 
an opportunity to present this talk.
This work is supported in part by Grant-in-Aid for Scientific Research
(A)\#12740144.

\end{document}